\def\bfseries{\fontseries\bfdefault\selectfont\boldmath}
\documentclass{ws-procs9x6}
\usepackage{graphics}
\usepackage{citesort}
\makeatletter
\@ifundefined{pdfoutput}%
        {\DeclareGraphicsExtensions{.eps,.ps}}%
        {\DeclareGraphicsExtensions{.pdf}}
\makeatother

\begin{document}

\title{Energy-Calibration of the ATLAS Hadronic and Electromagnetic 
       Liquid-Argon Endcap Calorimeters}

\author{Sven Menke\footnote{on behalf of the \uppercase{ATLAS} 
\uppercase{L}iquid \uppercase{A}rgon \uppercase{C}ollaboration}}

\address{Max-Planck-Institut f{\"u}r Physik\\
(Werner-Heisenberg-Institut)\\
F{\"o}hringer Ring 6,
80805 M{\"u}nchen, Germany\\
E-mail: menke@mppmu.mpg.de}


\maketitle

\abstracts{ In 2002 the first combined beam test of the hadronic and
electromagnetic liquid-argon endcap calorimeters of the ATLAS
experiment took place at the SPS test beam at CERN. A total of 15
million events from electrons, muons and pions in the energy range
from $6$ to $200\,{\rm GeV}$ were recorded. The entire calibration
chain, from digital filter weights, over calibration constants, to
clustering and energy weights, as is relevant for the energy
calibration of hadronic and electromagnetic showers in ATLAS was
tested and applied to the beam test data. The calibration methods and
first results for the combined performance of the two calorimeters are
presented.}

\section{Introduction}
\subsection{The ATLAS Hadronic and Electromagnetic Endcap Calorimeters}%
\label{subsec:calo}
The ATLAS hadronic (HEC) and electromagnetic (EMEC) liquid-argon
endcap
calorimeters~\cite{art:ATLAS-TDR96,art:ATLAS-Perf96,art:ATLAS-TDR99}
have to provide accurate measurements of jet energies and directions,
missing transverse energy and particle identification in the
pseudo-rapidity range $1.5 \le |\eta| \le 3.2$.  Methods relevant for
the energy calibration in ATLAS and first results of the combined
performance are subject of this presentation.
 
\subsection{The 2002 Beam Test Setup}\label{subsec:setup}
$1/8$ wheel of the EMEC was placed in front of 3 out of 32 HEC1
modules and 2 out of 32 HEC2 half-modules in a cryostat in the H6 test
beam area at the CERN SPS. The restriction to 2 half modules of the
second HEC wheel was mainly due to space constraints given by the
cryostat.  The 2 HEC2 half-modules were furthermore rotated by half a
$\phi$-segment around the beam axis w.r.t the nominal position in
ATLAS in order to minimize the leakage of hadronic energy.  In front
of the first EMEC layer a presampler endcap module placed inside the
cryostat allowed studies of preshower corrections with optional
additional material in front of the cryostat.

The beam particles hit the calorimeters at an angle of $90^\circ$ in
the region corresponding to $|\eta| \simeq 1.5-2.0$. Since both
calorimeters have readout structures pointing in $\eta$ the observed
response to electrons, pions, and muons was spread over more cells in
$\eta$ than expected in ATLAS.

Scintillators for triggering and timing and 4 MWPCs with horizontal
and vertical wire planes for beam position reconstruction were present
further upstream in the beam line.

\section{Signal Reconstruction}
The signal reconstruction follows closely the methods deployed in
previous stand-alone beam tests of the EMEC~\cite{art:EMECNIM} and the
HEC~\cite{art:HECNIM}.

The output from the EMEC and from the HEC summing amplifiers were
processed outside the cryostat in front-end-boards (FEB), which
perform the amplification of the EMEC signals and the signal shaping
for both calorimeters. A switched capacity array holds the digitized
samples at a sampling rate of $40\,{\rm MHz}$.

For each event 7 (16) samples per EMEC (HEC) channel were recorded
together with the MWPC response, trigger information and the TDC
measured delay between the trigger and the $40\,{\rm MHz}$ sampling
clock.

\subsection{Optimal Filtering}\label{subsec:OF}
The raw ADC samples are processed with an optimal filtering (OF)
 method~\cite{art:OF} using 5 samples.

For the HEC the detailed knowledge of each component in the
electronics chain and the form of the input calibration pulse is used
to determine the response function, which in turn is used to predict
the shape for physics signals.

For the EMEC with its more complicated electronic chain a numerical
method~\cite{art:ATL-LARG-2001-008} can be used to find the signal
shape from the measured calibration response and the Fourier
transformations of the ionization and the calibration current.

The resulting predicted physics shapes together with the
autocorrelation matrices from noise runs are used for the computation
of the OF weights. Unlike the final situation in ATLAS with its fixed
delay between trigger and sampling clock the beam test trigger comes
asynchronous w.r.t. the sampling clock. Therefore the OF weights are
calculated in steps of $0.5\,{\rm ns}$ in order to fill the $25\,{\rm
ns}$ trigger window and parameterized by a 4$^{\rm th}$ order
polynomial.

The achieved accuracy for the amplitude reconstruction following this
method is better than $1.5\,\%$ ($2\,\%$) for the HEC (EMEC).  The OF
reduces the noise of the amplitude to $64\,\%$ ($72\,\%$) of its
non-filtered value in the HEC (EMEC).

DAC level scans are used to find the conversion factors from ADC
counts to nA. The linearity is found to be better than $0.5\,\%$.

\section{Energy Reconstruction}
\subsection{Signal Corrections}\label{subsec:corrections}
The non-uniformity of the E-field and sampling-fraction variations
along the azimuth ($\phi$) in the EMEC are accounted for by applying a
correction (up to $2\,\%$) derived from the signal variation found for
electrons as a function of the reconstructed position in units of the
cell width in the second EMEC layer. Another correction (up to
$1\,\%$) is applied to account for residual signal variations with the
delay time between trigger and sampling clock. A potentially relevant
variation with $\eta$ is ignored due to the narrow $\eta$-range
considered in the beam test.

For the HEC a high voltage failure in 1 out of 16 LAr-gaps in the
second sampling for the middle module in $\phi$ makes a correction of
up to $15\,\%$ in this $\phi$-region in the 2$^{\rm nd}$ layer
necessary.

Following these corrections good uniformity for both calorimeters is
observed.

\subsection{Clustering}\label{subsec:cluster}
In each sampling a two-dimensional topological cluster algorithm is
used to define the group of readout cells relevant for analysis. Each
cluster consists of at least one cell with a signal-to-noise ratio
above 4 ($E > 4\sigma$). A threshold on the absolute value of the
signal-to-noise ratio, $|E| > 2\sigma$, is applied to all other
cells. They are included in the cluster if they share at least one
edge with a cluster member cell satisfying $|E| > 3\sigma$.  The
symmetric cuts on the cell and neighbor level avoid biases due to
electronics noise. Two super-clusters for the EMEC and the HEC are
defined by summing all cluster signals in the EMEC and the HEC,
respectively.  For the HEC the signals in the 3$^{\rm rd}$ layer are
multiplied by 2 in order to account for the $50\,\%$ smaller sampling
fraction.

\subsection{Response to Electrons}\label{subsec:electrons}
From Monte Carlo simulations of the test beam configuration the
leakage outside the EMEC was found to be very small for electrons and
is neglected.  Therefore the ratio of the known beam energy
($6-150\,{\rm GeV}$) and the sum of all signals in the EMEC in nA
defines the electromagnetic scale factor, $\alpha_{\rm em}^{\rm EMEC}
= 0.3855\pm0.004\,{\rm MeV}/{\rm nA}$, where the error is statistical
only. The variation with energy tests the linearity in the energy
range considered and was found to be better than $0.5\,\%$.

The energy resolution for electrons is studied with the super-cluster
in the EMEC which contains $96-98\,\%$ of the signal for high
energies.  Below $30\,{\rm GeV}$ the containment falls from $96\,\%$\
to $90\,\%$ for the lowest beam energies.

In data the resolution is found to be $\sigma_E/E =
(0.111\pm0.002)/\sqrt{E/{\rm GeV}} \oplus 0\pm0.001$ after noise
subtraction.  The noise $\sigma_{\rm noise} \simeq 0.2-0.3\,{\rm GeV}$
varies with energy due to the non-fixed cluster
size. Geant3~\cite{art:Geant3,art:Geant3HEC}
(Geant4~\cite{art:Geant4,art:HPW}) based Monte Carlo simulations yield
slightly better (worse) resolution results.

\subsection{Response to Pions}\label{subsec:pions}
The electromagnetic scale for the HEC is taken from the previous
stand-alone beam test~\cite{art:HECNIM}, $\alpha_{\rm em}^{\rm HEC} =
3.27\pm0.03\,{\rm MeV}/{\rm nA}$, taking the modified electronics into
account.  Good agreement of the total visible energy in the EMEC and
HEC for pions with Monte Carlo simulations based either on Geant3 or
the quark-string-gluon-plasma (QGSP) model of Geant4 is observed while
the Geant4 low-and-high-energy-pion-parameterization (LHEP) model
deviates largely from data.

\subsection{Weighting}\label{subsec:weighting}
The non-compensating nature of the two calorimeters makes weighting of
hadronic energy deposits necessary. A cell based weighting method
which was successfully used in previous
experiments~\cite{art:H1det,art:H1larg} needs a detailed simulation on
the cell level, which is not yet available for ATLAS.  Therefore a
more coarse weighting scheme on the super-cluster level has been
applied.

With the leakage outside the detector volume as predicted by the Monte
Carlo and the known beam energy 6 weights (3 for the EMEC and 3 for
the HEC) are fitted from the two super-cluster energies and their
energy density, leading to the weighted energies $E_{\rm w} = E_{\rm
em}\left(C_1\cdot\exp\left[-C_2 \cdot E_{\rm em}/V\right]+C_3\right)$.

The noise subtracted resolution for negative pions $\sigma_E/E =
(0.827\pm0.003)/\sqrt{E/{\rm GeV}} \oplus 0\pm0.003$ is slightly worse
than expected from Monte Carlo, which gives sampling terms around
$70\,\%$. For positive pions a sampling term of $(79.9\pm0.4)\%$ is
found in data.

The ratio of the combined weighted energy of EMEC and HEC over the
combined electromagnetic energy yields the effective ${\rm
e}/\pi$-ratio for the endcap calorimeters ranging from $1.32$ at
$20\,{\rm GeV}$ to $1.19$ at $200\,{\rm GeV}$ for negative pions and
$0.05$ larger values for positive pions. Geant4 based simulations
predict smaller ratios by $0.02$ for LHEP and $0.05$ for QGSP, while
Geant3 is off by $-0.09$.


\end{document}